\documentclass[prd,preprint,groupedaddress,nofootinbib,showpacs,preprintnumbers,11pt]{revtex4}

\pdfoutput=1
\usepackage{hyperref}

\usepackage{amssymb}
\usepackage{amsbsy}
\usepackage{amsmath}
\usepackage{url}
\newcommand{\rf}[1]{(\ref{#1})}
\newcommand{\beq}{\begin{equation}}
\newcommand{\eeq}{\end{equation}}
\newcommand{\be}{\begin{equation}}
\newcommand{\ee}{\end{equation}}
\newcommand{\bea}{\begin{eqnarray}}
\newcommand{\eea}{\end{eqnarray}}
\newcommand{\eq}[1]{Eq.~(\ref{#1})}
\newcommand{\non}{\nonumber \\*}

\newcommand{\bv}{\begin{verbatim}}
\newcommand{\ev}{\end{verbatim}}

\newcommand{\g}{u}
\newcommand{\tbeta}{\tilde\beta}
\newcommand{\vp}{\phi}

\newcommand{\e}{\,\mbox{e}}
\renewcommand{\d}{{\rm d}}
\renewcommand{\i}{{\rm i}}
\newcommand{\blambda}{\bar\lambda}
\newcommand{\brho}{\bar\rho}

\newcommand{\C}{\blambda}

\newcommand{\hg}{\hat{g}}

\newcommand{\om}{\omega}

\def\fun#1#2{\lower3.6pt\vbox{\baselineskip0pt\lineskip.9pt
\ialign{$\mathsurround=0pt#1\hfil##\hfil$\crcr#2\crcr\sim\crcr}}}

\begin{document}


\title{The susceptibility exponent of Nambu-Goto strings}

\author
{Jan Ambj\o rn$\,^{a,b}$ and Yuri Makeenko$\,^{a,c}$}

\affiliation{\vspace*{2mm}
${}^a$\/The Niels Bohr Institute, Copenhagen University,
Blegdamsvej 17, DK-2100 Copenhagen, Denmark\\
${}^b$\/IMAPP, Radboud University, Heyendaalseweg 135,\\
6525 AJ, Nijmegen, The Netherlands\\
${}^c$\/Institute of Theoretical and Experimental Physics,
B. Cheremushkinskaya 25, 117218 Moscow, Russia\\
\vspace*{1mm}
{email: ambjorn@nbi.dk \ makeenko@nbi.dk}
}


\begin{abstract}

We compute the string susceptibility $\gamma_{\rm str}$ for 
the regularized Nambu-Goto string in $d$ dimensions and  obtain  $\gamma_{\rm str}=1/2 $  in $2<d<26$.
This  agrees with  previous results obtained for lattice strings. \\

\end{abstract}


\maketitle

\section{Introduction}

An important characteristics of string dynamics is the string susceptibility 
$\gamma_{\rm str}$ which is determined by the pre-exponential in the entropy of surfaces with
 large area. For the Polyakov string of genus $g$ embedded in 
$d$-dimensional space-time it equals
the Knizhnik-Polyakov-Zamolodchikov--David--Distler-Kawai
(KPZ-DDK) value~\cite{KPZ}  
\beq
\gamma_{\rm str}^{(g)} = \gamma_{\rm str}^{(0)} + g \big( 2- \gamma_{\rm str}^{(0)}\big),\qquad 
\gamma_{\rm str}^{(0)} = \frac{c-1- \sqrt{(25-c)(1-c)}}{12}, 
\label{KPZ}
\eeq
where $c=d-1 \leq 1$ is the central charge  of conformal matter.
Equation~\rf{KPZ} beautifully describes a vast amount of the models with $c\leq1$ and it follows in a simple way
from the conformal invariance of the quantum Liouville action, as we will briefly review in Sec.\ \ref{s:gamma}, 
but it apparently breaks down for $1<c<25$ (or $2<d<26$), where 
the right-hand side of \rf{KPZ} becomes complex. 

The non-trivial dynamical information about the coupling of geometry and matter 
is present already in $\gamma_{\rm str}^{(0)}$, i.e.\ for genus 0, as is apparent from \rf{KPZ}. On the other hand, $g=1$ has 
in this respect a special status since $\gamma_{\rm str}^{(1)}=2$, independent of the coupling of surface geometry to matter, 
reflecting that the integral of  the intrinsic curvature for a closed surface of genus $g=1$ is zero. 

Models where $c> 1~(d > 2)$ have been addressed for strings with an explicit UV cutoff and 
 for  the Nambu-Goto string on a hypercubic lattice it was obtained \cite{lattice} (see also \cite{ADJ97} and references therein):
\be
\gamma^{(0)}_{\rm str} =\frac12 \qquad \hbox{for $d>2$} 
\label{1/2}
\ee
which is a typical behavior for branched polymers. The same value \rf{1/2} has been
obtained~\cite{AM17a,AM17c} for the Nambu-Goto string with the proper-time
or Pauli-Villars regularizations at $2<d<26$ in the mean-field approximation. Further, it has been argued~\cite{bergfin} that 
the left equation in \rf{KPZ} is valid even for models with $c > 1$, i.e. in particular that $\gamma_{\rm str}^{(1)}=2$ even 
for these models.

The purpose of this Letter is twofold. Firstly, we repeat the computation of 
$\gamma^{(0)}_{\rm str}$ of \cite{AM17c} for closed surfaces, such that one gets rid of the boundary effect 
present in \cite{AM17a,AM17c} where strings with the topology of a cylinder were considered. We do that 
by compactifying target space $R^d$ to $ T^2\times R^{d-2}$, where $T^2$ is a torus  of fixed periods $\beta$ and $L$. 
Secondly, we go beyond the mean-field approximation and show that \eq{1/2} is {\em exact}\/ for $2<d<26$.

\section{The setup for a torus\label{s:torus}}

The results~\cite{AM17a,AM17c} for the long cylinder
of length $L$ and circumference $\beta$ are expected to
be analogous to those for the long torus modulo boundary effects which are suppressed
as $\beta/L$ for the cylinder. Let us demonstrate this by explicit computations for the torus.

The setup is the following: we consider the closed bosonic string  in $d$-dimensional target space, where direction 1 and 2  
are compactified to a torus, $S^1\times S^1$, with periods $\beta$ and $L$. We use the Nambu-Goto action
\beq\label{j1}
S_{NG} [X^\mu] = K_0 \int \d^2\om \; \sqrt{\det \partial_a X^\mu \partial_b X^\mu},
\eeq
where  $K_0$ is the bare string tension. 
In addition we insist that the string winds around the target-space  torus. A classical solution, minimizing the action under the 
condition that it winds around the target-space torus, is then
 \be
X^1_{\rm cl}= \beta \om_1 , \quad X^2_{\rm cl}=L \om_2,\qquad \rho^{\rm cl}_{ab} = 
\partial_a X_{\rm cl}^\mu \,\partial_b X_{\rm cl}^\mu
=\begin{pmatrix}\beta^2 & 0 \\ 0 & L^2 \end{pmatrix},   \qquad  S_{NG} [X_{\rm cl}^\mu] = K_0  L \beta,
\label{ja2}
\ee
where the parameters $\om_1,\om_2\in [0,1]$ also belong to a torus with periods 1.

In order to use the Nambu-Goto action in the path integral we follow the setup \cite{AM17a} for a long cylinder and introduce the Lagrange multiplier $\lambda^{ab}$ and the intrinsic metric tensor $\rho_{ab}$ such that 
\beq\label{j3}
S_{NG}(X^\mu, \lambda^{ab},\rho_{ab}) = K_0 \int \d^2 \om \; \sqrt{\det \rho_{ab}} + \frac{K_0}{2} \int \d^2 \om \, \lambda^{ab} 
\Big(  \partial_a X^\mu   \partial_b X^\mu -\rho_{ab}\Big).
\eeq
The path integral now involves the integration over $X^\mu$, $\rho_{ab}$ and $\lambda^{ab}$. The integration over the $X^\mu$ 
is performed by writing $X^\mu = X^\mu_{\rm cl} + X^\mu_{\rm q}$ and performing the Gaussian integration over the quantum fluctuations $X^\mu_{\rm q}$. 
The integration over $\rho_{ab}$ requires a gauge fixing and we choose conformal gauge. We can then write 
\beq\label{ja4}
\rho_{ab} (\omega) = \rho(\omega) \, \hg_{ab}(\tau), \qquad \hg_{ab} (\tau) = 
\begin{pmatrix} 1 & \tau_1 \\ \tau_1 & \tau_1^2 + \tau_2^2 \end{pmatrix}, \qquad  \hg = \det \hg_{ab} = \tau_2^2,
\eeq
where $\tau = \tau_1 +i \tau_2$ is the modular parameter for the torus. For our choice of the classical solution \rf{ja2} we have
\beq\label{ja4a}
\rho_{ab}^{\rm cl} = \rho^{\rm cl} \hg_{ab} (\tau^{\rm cl}), \qquad  \rho^{\rm cl} = \beta^2, 
\quad  \tau^{\rm cl}_1=0,\quad \tau^{\rm cl}_2= \frac{L}{\beta}, \quad \sqrt{\det \rho_{ab}^{\rm cl}} = L \beta .
\eeq 
The integration over $X^\mu_q$ will result in a term \cite{AM17a,AM17c}
\beq\label{ja5}
\Big( \det \, {\cal O} \Big)^{-d/2},\qquad {\cal O} = - \frac{1}{\sqrt{\hg}}\, \partial_a \lambda^{ab} \partial_b
\eeq
and similarly there is a ghost term $\det ({\cal O}_{\rm gh})$ from choosing the conformal gauge \rf{ja4}. 
Including the determinants in 
an effective action $S_{\rm eff}[\lambda^{ab},\,\rho]$ and adding a source term, 
we can write
\be
Z[J]= \int  
 \frac{\d\tau_1\d \tau_2}{\tau_2}\int {\cal D} \lambda^{ab} \int {\cal D} \rho \, 
\e^{-S_{\rm eff}[\lambda^{ab},\,\rho,J]} , \quad
S_{\rm eff}[\lambda^{ab},\,\rho,J]=S_{\rm eff}[\lambda^{ab},\,\rho]+
\frac 12 \int \sqrt{\hat g}\hat g_{ab}J^{ab}  \rho.
\label{Z[J]}
\ee 

In the mean-field approximation advocated in \cite{AM17a} we have that $\rho_{ab}(\om)$ and $\lambda^{ab}(\om)$ are 
 independent of $\om$ for  the worldsheet coordinates we use.  We can thus write 
 \beq\label{ja6}
\bar{ \rho}_{ab} = \bar{\rho} \, \hg_{ab}(\tau), \quad \sqrt{\det \bar{\rho}_{ab}} = \bar{\rho} \tau_2, \qquad \quad
 \blambda^{ab} = \bar{\lambda} \,\sqrt{\hg} \, \hg^{ab}(\tau)=\blambda \tau_2\,\hg^{ab}(\tau), 
 \quad \sqrt{\det \blambda^{ab}} = {\bar{\lambda}},
 \eeq
 where $\bar{\rho}$ and $\bar{\lambda}$ are constants. In this case the determininants can be calculated\footnote{The cutoff-independent finite part of $\ln \det ( -\hat g^{ab} \partial_a \partial_b)$ is given by $- \pi \tau_2 /3 +F(\tau)$ where $F(\tau)/\tau_2 \to 0$ for 
 $\tau_2 \to \infty$~\cite{Pol86}.} and we obtain in the limit where $L \gg \beta$, using the Pauli-Villars regularization described 
 in \cite{AM17c}: 
 \bea
\lefteqn{S_{\rm eff}[\blambda^{ab},\brho,J]} \non
&&= 
\frac {K_0}2  \blambda^{ab} \Big(\rho^{\rm cl}_{ab}
-\brho_{ab}\Big)   
+\left[
K_0+J-\left(\frac{d}{2 \sqrt{\det \blambda^{ab}}}-1 \right)
 \Lambda^2\right] \sqrt{\det \brho_{ab}} 
-\frac{\pi(d-2) }{6} 
\frac{\sqrt{\det \blambda^{ab}}}{\lambda^{22}}
\nonumber\\
&&=
\frac{K_0 \blambda}{2\tau_2} \Big[
 (\tau_1^2+\tau_2^2){\beta^2}
+{L^2}\Big]
+\left[K_0 (1-\blambda)-\frac{d\Lambda^2}{2 \blambda}+\Lambda^2+J\right]\tau_2
 \brho
-\frac{\pi(d-2)\tau_2 }{6} ,
\label{B4p}
\eea
where  $\Lambda$ is a UV cutoff.

Minimizing \rf{B4p} with respect to $\brho$, we find
\be
\blambda(J)=\frac{1}2  \left(1+\frac J{K_0}+\frac{\Lambda^2}{K_0} \right) +\frac 12
\sqrt{\left(1 +\frac J{K_0}+\frac{\Lambda^2}{K_0}\right)^2 -\frac{2d\Lambda^2}{K_0}}
\label{newC}
\ee
which is associated with the mean-field approximation. As was shown in~\cite{AM17a},
this minimum  in $2<d<26$ is favorable to the usual classical minimum and stable under
local fluctuations.
At this value of $\blambda$ the coefficient in front of $\brho$ in  \eq{B4p}
vanishes so this term vanishes at the minimum and we find
\bea
S_{\rm eff}[\blambda, \brho,J]&=&\frac{K_0 \blambda}{2\tau_2} \left[
 (\tau_1^2+\tau_2^2) {\beta^2}
+{L^2}\right]
-\frac{\pi(d-2)\tau_2 }{6} .
\label{B4pp}
\eea

As far as the integral over the modular parameters is concerned, it has a saddle point 
for large $L$.
The saddle point occurs at the values 
\be
\bar \tau_1= 0, \qquad
\bar \tau_2 =\frac{L}{\sqrt{\beta^2-\frac{\pi(d-2)}{3 K_0 \blambda(J)}}}.
\label{sadpoi}
\ee
For these values we have 
\beq\label{ja10}
 \brho  =   
\frac { \C(J)} {\sqrt{ \left(1 +\frac J{K_0}+\frac{\Lambda^2}{K_0}\right)^2-\frac{2d\Lambda^2}{K_0}}}\;
\Big[\beta^2-\frac{\pi(d-2)}{6 K_0 \blambda(J)}\Big].
\eeq
Finally the value of the effective action in the mean-field approximation for $L\gg \beta$  reads:
\be
S_{\rm eff}[\blambda, \brho,J]=K_0 \blambda(J)\, L\,\sqrt{\beta^2-\frac{\pi(d-2)}{3 K_0 \blambda(J)}}.
\ee

Notice that for long strings with  $L \gg \beta \gg 1\sqrt{K_0}$ the values \rf{sadpoi} of the modular parameters are close to 
the values \rf{ja4a} for the classical induced metric $\rho_{ab}^{\rm cl}$, 
and in fact  $\brho_{ab}$ becomes 
proportional to the classical value \rf{ja4a}:
\be
\brho_{ab} (J)=
\frac { \C(J)} {\sqrt{ \left(1 +\frac J{K_0}+\frac{\Lambda^2}{K_0}\right)^2-\frac{2d\Lambda^2}{K_0}}} \; \rho_{ab}^{\rm cl} + O(\beta/K_0).
\label{newrho}
\ee

We finally note that for $J=0$ we can introduce renormalized  target-space lengths $L_R$, $\beta_R$ and a renormalized 
coupling constant (string tension) $K_R$ by 
\beq\label{ja11} 
L_R = \sqrt{ \frac{\blambda(0)}{2\blambda(0) - 1-\frac{\Lambda^2}{K_0}}} \;L,\quad
\beta_R = \sqrt{ \frac{\blambda(0)}{2\blambda(0) - 1-\frac{\Lambda^2}{K_0}}} \; \beta,\quad
K_R =\Big( 2\blambda(0) - 1-\frac{\Lambda^2}{K_0}\Big)\; K_0,
\eeq
such that the expressions for $S_{\rm eff}$ and   $\brho_{ab}$  stay finite when the cutoff $\Lambda \to \infty$.

Thus we have explicitly  demonstrated that  the results for
the long cylinder and the long torus are  analogous modulo boundary effects which are suppressed 
for the cylinder as $\beta/L$. 

\section{Mean-field approximation for string susceptibility
\label{s:gamma}}

The string susceptibility $\gamma_{\rm str}$ characterizes the string entropy.
In the case of Liouville gravity we have for $c < 1$, after integration out the matter fields, a partition function for 
surfaces of genus $g$:
\beq\label{ja20}
Z(\mu) = \int \d m(\tau_i) \int  {\cal D} \phi \; \e^{-S_L [\phi,\hat{g}]},\qquad 
S_L[\phi,\hat{g}] = \frac{1}{16\pi b^2} \int \d^2 \om \sqrt{\hat{g}}\; 
\big(\partial_a \phi \partial^a \phi + 2b Q \hat R \phi + b^2\mu \e^{\phi}\big)
\eeq
where $b+1/b = Q = \sqrt{(25-c)/6}$,
$\hat R$ is the scalar curvature of the metric $\hat g_{ab}$
and where $\d m(\tau_i)$ denotes the integration over the modular parameters
$\tau_i$ for surfaces of genus $g$. This is the partition function for an ensemble of surfaces with cosmological constant $\mu$.
The partition function $Z(A)$ of a canonical ensemble of surfaces with fixed area $A$ is related to $Z(\mu)$ by a Laplace 
transformation 
\beq\label{ja21}
Z(\mu) = \int_0^\infty \d A \; \e^{-\mu A}\, Z(A), 
\eeq
\beq\label{ja22}
Z(A) =  \int  \d m(\tau_i)\int  {\cal D} \phi \, \e^{ -\frac{1}{16\pi b^2} \int \d^2 \om \sqrt{\hat{g}}\, 
\big(\partial_a \phi \partial^a \phi + 2b Q  \hat R \phi \big)} \; \delta \left(  \int d^2\om \, \sqrt{\hat{g}}\,\e^{\phi} - A\right).
\eeq

By using the fact that a scaling of $A$ can be compensated by a shift $\phi \to \phi +c$ one arrives at 
\beq\label{ja23}
Z(A) = A^{(g-1)Q/2b-1} \  Z(A=1) := A^{\gamma_{\rm str}^{(g)}-3} Z(A=1).
\eeq
With this definition of $\gamma_{\rm str}^{(g)}-3$ one obtains \rf{KPZ}. We remark the following: firstly, it is clear that 
$\gamma_{\rm str}^{(1)}=2$ and that this scaling comes entirely from the $\delta$-function in \rf{ja22}. Secondly, we see that 
the scaling \rf{ja23} formally translates into a scaling 
\beq\label{ja24}
Z(\mu) \propto \mu^{2- \gamma_{\rm str}^{(g)}}, \quad {\rm i.e.} \quad 
\frac{ \d^2 Z(\mu)}{ \d \mu^2} = \frac{{\rm const}}{\mu^{\gamma_{\rm str}^{(g)}}}
\eeq 
which is the reason for the notation ``string susceptibility'', by analogy with the susceptibilty of a spin-system, and the origin
of the $-\!3$ in $\gamma_{\rm str}\!-\!3$. Thirdly, the relation to the entropy of the number of surfaces is particularly transparent if 
we use the Nambu-Goto action. In this case  we have formally 
\beq\label{ja25}
Z(\mu) = \int  {\cal D} S\; \e^{-\mu  A(S)} =  \int_0^\infty \d A \; \e^{-\mu A} Z(A),\qquad Z(A) = \int  {\cal D} S \; \delta( A(S) -A),
\eeq
where the integration is over embedded surfaces $S$ in the target space. In this case $Z(A)$ is just the formal number of 
embedded surfaces. 

This number is of course infinite. In \rf{ja23} this infinity has been renormalized away, but in a quantum 
theory where we keep the UV cutoff it will appear as an exponential growing number depending on the cutoff, but remarkably,
and important for consistency of string theory, independent of the genus $g$ of the surfaces. Thus in general we will write 
\beq\label{ja26}
Z(A) \propto  A^{\gamma_{\rm str}^{(g)}-3} \; \e^{C(\Lambda) A}(1 + O(1/A)),
\eeq
where $C(\Lambda)$ is a cutoff-dependent constant which can also depend on some of the other bare coupling constants of the model, and where the correction indicates that in a regularized theory we 
only expect the formula to be correct for areas much larger than $1/\Lambda^2$ and also much larger than any dimensionful bare
coupling constant in the appropriate power.  

Until now we have been considering closed surfaces. Consider marking a point on the surface. For such a marked 
surface one would expect 
\beq\label{ja27}
Z_m (A) \propto   A^{\gamma_{\rm str}^{(g)}-2} \; \e^{C(\Lambda) A}(1 + O(1/A)),
\eeq
simply because we can put the mark anywhere on the surface, and that should produce a factor proportional to $A$.
External length-scales can be introduced either by considering vertex operators producing generalized marked points
or by considering for instance strings fixed at a boundary in target space. 
Topologically, having such a boundary, say a rectangular planar loop with side lengths $L$ and $\beta$ in target space, 
corresponds to a surface with a marked point, and using such surfaces for the calculation of  $\gamma_{\rm str}^{(g)}$
one should use formula \rf{ja27}, since for $A \gg A_{\rm min} = L\beta$ the boundary looks essentially like a point. A similar
argument applies to our present setup. Here we are not changing the topology of the surface, but we are forcing it to wind around
a torus in target space which has area $A_{\rm min} = L \beta$, which then acts much like fixing a boundary. But in addition we 
should also change $g \to g-1$ in \rf{ja27}, since we have already, by the explicit setup,  used up one of the ``handles'' $g$ 
going around the small target space torus. The large area fluctuations which contributes to the entropy will thus be fluctuations 
corresponding to genus $g-1$. In our case we consider surfaces of $g=1$ and by our setup, finding the large $A$ behavior of 
$Z(A)$, using formula \rf{ja27}, we will therefore determine $\gamma_{\rm str}^{(0)}$.

Thus we consider the following fixed area partition function
\be
 Z(A)=\int  
 \frac{\d \tau_1\d \tau_2}{\tau_2}\int {\cal D} \lambda^{ab} \int
{\cal D} \rho \, \e^{-S_{\rm eff}[\lambda^{ab},\,\rho]}\,
\delta \Big( \int \d^2 \om\,\rho - A \Big) \stackrel{A\to\infty}\propto A^{\gamma_{\rm str}^{(0)}-2} \e^{C A} .
\label{ggstr}
\ee
However in order to use our mean-field results we want to get rid of the $\delta$-function in \rf{ggstr} and we 
do that by a  Lagrange multiplier $J$ (i.e.\ in the notation of \rf{ja21} we express $Z(A)$ as the inverse  Laplace transform of $Z(\mu)$):
\be
Z(A)= \int_\uparrow \frac{\d J}{2\pi\i}  
\int  
 \frac{\d \tau_1\d \tau_2}{\tau_2}\int {\cal D} \lambda^{ab}
 \int {\cal D} \rho \,\e^{-S_{\rm eff}[\lambda^{ab},\rho,J]+J A} ,
\label{FFF}
\ee
where the integral over $J$  runs along the imaginary axis. 

In the mean-field approximation we compute
the integral over $J$  in \rf{FFF},
expanding the exponent   about the saddle point  at 
\be
\bar J(A)=\sqrt{{2d\Lambda^2 K_0}}\frac{(A-A_{\rm min}/2)}{ \sqrt{A(A-A_{\rm min})}}
-K_0-{\Lambda^2},\qquad A_{\rm min}=L\beta.
\label{jjrr}
\ee
To quadratic order in  $\Delta J=J-\bar J(A)$  we find for the exponent in \rf{FFF} 
\be
 {JA}- \C(J)  {K_0A_{\rm min}}= \sqrt{{2d \Lambda^2}{K_0}} \sqrt{A(A-A_{\rm min})}
- \left(K_0+{\Lambda^2} \right) A
+\sqrt{\frac{2}{d\Lambda^2 K_0}}\frac{A^3}{A_{\rm min}^2}
\left(\Delta J\right)^2 +\ldots\,.
\label{15}
\ee
Integrating over $\Delta J$  along the imaginary axis 
and over the modular parameters $\tau_1$ and $\tau_2$ about
the saddle point \rf{sadpoi},  we finally obtain
\be
\log Z(A)=-\sqrt{{2d \Lambda^2}{K_0}} \sqrt{A(A-A_{\rm min})}+ \left( K_0+\Lambda^2\right)A-
\frac3{2 } \log   A  +\frac 12 \log {A_{\rm min}} +{\rm const.}
\label{16}
\ee
Comparing with the definition \rf{ggstr}, this gives~\cite{AM17c} 
$\gamma_{\rm str}^{(0)}=1/2$
in the mean-field approximation.

Note also that we indeed obtain the announced form \rf{ja27} for $Z(A)$:
\beq\label{ja28}
Z(A) \propto A^{-3/2} \e^{C(\Lambda,K_0)\; A} (1+ O(1/A)), \qquad C(\Lambda,K_0) = K_0 +\Lambda^2 - \sqrt{2d \Lambda^2 K_0}.
\eeq
As discussed in \cite{AM17a,AM17c} the so-called critical point $K_0 =K_*$, where scale invariance should be restored 
when $\Lambda \to \infty$, is precisely the point where $C(\Lambda,K_0) =0$,
\beq\label{ja29}
K_* = [(d-1) + \sqrt{d(d-2)}] \, \Lambda^2,
\eeq 
which is also the point where the constant of proportionality between $\brho_{ab}$ and $\rho_{ab}^{\rm cl}$ in \rf{newrho}
diverges since $2 \lambda(0) K_0 -(K_0+ \Lambda^2) = \sqrt{(K_0 +\Lambda^2)^2-2d\Lambda^2K_0}$ according to 
\rf{newC}. To obtain a finite result for $\brho$ one has to choose $K_0$ infinitesimally larger than $K_*$, as 
\beq\label{ja30}
K_0= K_* +\frac{K_R^2}{2\Lambda^2 \sqrt{d(d-2)}}+ O(K_R^2/\Lambda^6),
\eeq
which is precisely the renormalization of $K_0$ given in \rf{ja11}.

\section{Beyond the mean field}

To account for fluctuations about the mean field, we write $\rho=\brho \e^\vp$ 
and compute the effective action for $\brho$, the slow part of the metric, 
 by averaging over $\vp$ associated with the fast part.
An important observation is that we need only the divergent part of the effective action 
because the finite part does not affect $\gamma_{\rm str}$.
This divergent part comes only from tadpole diagrams. 

We perform
the computation of  (the divergent part of)  the effective action  \rf{B4p}
by path-integrating over $\vp$. The result has  the form
\bea
S_{\rm eff}^{\rm div}[\lambda^{ab},\brho,J]
&=&\int  \d^2 \om \left\{
\frac {K_0}2 
\Big[\lambda^{11}  {\beta^2}
+ \lambda^{22} {L^2}
-\left[\lambda^{11}+2\lambda^{12} \tau_1+
\lambda^{22}(\tau_1^2+\tau_2^2)\right] \brho\Big]\right. \non
&& \hspace*{1cm}\left. +\left[
K_0+J-\left(\frac{d}{2 \sqrt{\det \lambda^{ab}}}f_1(b^2)-f_2(b^2) \right)
 \Lambda^2\right] \tau_2 \brho \right\} 
\label{Sdiv}
\eea
and involves two functions $f_1(b^2)$ and $f_2(b^2)$, coming respectively from the matter and ghost tadpoles, whose Taylor's expansions in $b^2$
(that multiplies the propagator of $\vp$ as usual in the Liouville field notation (see \rf{ja20}))
starting  from 1.
Since  the Lagrange multiplier
$\lambda^{ab} $ does not propagate, we can replace it for the torus
by a constant value
$\lambda^{ab} = \blambda \sqrt{\hat g}\, \hat g^{ab} $ 
and analogously  $\brho$ is constant for our choice of the coordinates.
Then the action \rf{Sdiv}   
 takes the form analogous to \rf{B4p}:
\be
S_{\rm eff}^{\rm div}[\blambda,\brho,J]
=\frac{K_0 \blambda}{2\tau_2} \Big[
 (\tau_1^2+\tau_2^2) {\beta^2}
+{L^2}\Big]
+\left[K_0(1-\blambda)+J-\left(\frac{d}{2\blambda}\Lambda_1^2-\Lambda^2_2 \right)
 \right]\tau_2 \brho , 
\label{Sdiv1rr}
\ee
where for latter convenience we have introduced  the notation 
\be
\Lambda_1^2=f_1(b^2)\Lambda^2, \qquad \Lambda_2^2=f_2(b^2)\Lambda^2.
\ee

The field $\brho$ is analogous to constant fields generated in the problems of
spontaneous symmetry breaking. It does not fluctuate if
the volume is large and can be substituted by its value minimizing the effective
action \rf{Sdiv1rr}.  We finally arrive at the values of $\blambda$ and $\brho$
\begin{subequations}
\bea
\blambda(J)&=&\frac{1}2  \left(1+\frac{J}{K_0}+\frac{\Lambda_2^2}{K_0} \right) +\frac 12
\sqrt{\left(1 +\frac{J}{K_0}+\frac{\Lambda^2_2}{K_0}\right)^2 -\frac{2d\Lambda^2_1}{K_0}},~~
\label{newC1} \\
\brho(J)&=&
\frac {\C(J)} {\sqrt{ \left(1 +\frac J{K_0}+\frac{\Lambda_2^2}{K_0}\right)^2 -\frac{2d\Lambda_1^2}{K_0}}}
 {\beta^2},
\label{newrhor1r}
\label{newlarr}
\eea
\label{mmff1}
\end{subequations}
generalizing  Eqs.~\rf{newC} and \rf{ja10} (for $\beta^2 \gg 1/K_0$).

Substituting in  \eq{FFF}, we obtain 
\be
Z(A)= \int  
 \frac{\d \tau_1\d \tau_2}{\tau_2}
\int_\uparrow \frac{\d J}{2\pi\i}   \,\e^{J A- \C(J) 
 {K_0L\tilde\beta}/2} ,
\label{FFFtor}
\ee
where
\be
\tilde \beta = \frac\beta{2\tau_2} {\left[(\tau_1^2+\tau_2^2)\frac \beta L 
+\frac{ L}{\beta} \right]}.
\label{tbeta}
\ee

Let us concentrate on the scaling limit  when $K_0$ approaches the critical value
\be
K_*=d\Lambda_1^2-\Lambda_2^2+\Lambda_1 \sqrt{d^2 \Lambda_1^2 -2d\Lambda_2^2}
\ee
as
\be
K_0
=
K_*+\frac{K_R^2}{2\Lambda_1\sqrt{d^2 \Lambda_1^2 -2d\Lambda_2^2}}.
\ee
This is a generalization of the scaling limit already discussed in Eqs.~\rf{ja29} and \rf{ja30}.
We then have either particle-like or string-like behavior in the scaling regime, in the terminology of \cite{AM17a}, and 
\bea
J A - S_{\rm eff} [\blambda,\brho,J]&= &JA -K_0 \blambda(J) L\tbeta  \non
&\stackrel{\rm s.l.}\to& J \left(A- \frac{L\tilde \beta }2\right) -\frac{1}{2} L\tbeta 
\sqrt{K_R^2+2 J(K_*+\Lambda_2^2)+J^2}-\frac {L\tbeta}2  ( K_*+\Lambda_2^2).
\label{88}
\eea
Only the domain $J\ll K_*$ will be essential in the integral over $J$ for $A\gg L\tilde\beta$
as is seen from 
\eq{jjrr}, so we can drop $J^2$ under the square root in \eq{88}.
Introducing the new variables%
\footnote{To avoid confusion 
let us note that $K_*+\Lambda_2^2=\sqrt{2d\Lambda_1^2 K_*}$.}
\be
\g= J A,\qquad
B=\frac{A K_R^2}{K_*+\Lambda^2_2},\qquad c=\frac{K_R L\tbeta}{2} ,
\ee
we then rewrite \rf{88} as
\be
\rf{88}=JA  
- \frac {L\tilde \beta}2\sqrt{ K_R^2  +2 J (K_*+\Lambda^2)   }=\g- c\sqrt{1+\frac  {2\g} B} .
\ee

These $B$ and $c$ are both large as $\sim K_R$ for large $K_R$, 
so the results of the previous section
for the one-loop order of the expansion about the mean field
can be reproduced, expanding the integral over $\g$ 
about the saddle point 
\be
\bar \g = \frac 12 \left(  \frac {c^2}B-B \right).
\ee
Then
\be
\g- c\sqrt{1+\frac  {2\g} B} = -\frac{B^2+c^2}{2B}+\frac{B}{2c^2} (\g-\bar \g)^2+\ldots.
\label{36}
\ee
This results in the distribution
\be
Z(A) 
= \int  
\frac{\d \tau_1\d \tau_2}{\tau_2}\e^{-\sqrt{2\Lambda_1^2K_*}L\tilde\beta/2 }
\frac{c}{A\sqrt{2\pi B}}\e^{- {B}-c^2/2 B},
\label{81}
\ee
which clearly shows the same dependence on $A$ as in \eq{16}.
The remaining in \eq{81} integrals over the modular parameters do not change the
$A$-dependence at large $A$ and only give the same normalization factor as in \eq{16}.
How to compute these integrals will be described at the end of this section.

We can continue the analysis, introducing the new integration variable
\be
x=\sqrt{B+2\g} - \frac c{\sqrt B}
\ee
to rewrite the integral over $J$ in \eq{FFFtor} as
\bea
\int_\uparrow \frac{\d \g}{2\pi\i A} 
  \,\e^{\g- c\sqrt{1+\frac  {2\g} B}} 
&=& \e^{- (B+c^2/B)/2}
\int \frac{\d x}{2\pi\i A}\left(x +\frac c{\sqrt B}\right) \e^{x^2/2} \non &=&
 \e^{-  (B+c^2/B)/2 } \left[
\frac{1}{2\pi \i A} \e^{x^2/2}\Big |_{x_-}^{x_+} +
\frac{c}{2\i A\sqrt{2\pi B} }{\rm Erfi}(x/\sqrt{2}) \Big|_{x_-}^{x_+} \right]
\eea
with
\be
x_\pm=\sqrt{B\pm2\i \infty} - \frac c{\sqrt B}.
\ee
This gives precisely  the distribution \rf{81} whose one-loop approximation 
calculated by the expansion \rf{36} is thus exact.

The remaining integrals over the modular parameters 
$\tau_1$ and $\tau_2$ in \eq{81} can be easily computed at large $L\beta$
by the saddle point
given by \eq{sadpoi} and which is justified by large 
$L\beta \sqrt{2d \Lambda_1^2 K_*}$. Accounting for the fluctuations about this saddle point,
the final result for the long torus 
 is 
\be
\log Z(A)=\frac {L\beta}2  \sqrt{2d\Lambda_1 ^2 K_*}-
\frac3{2 } \log   A  +\frac 12 \log A_{\rm min}+{\rm const.},\qquad A_{\rm min}={L\beta}
\label{16pp}
\ee
which reproduces the scaling regime of \rf{16} for $\Lambda_1=\Lambda$.
We have thus demonstrated that this behavior is {\em exact},  yielding 
$\gamma_{\rm str}^{(0)}=1/2$ for bosonic string in $2< d < 26$ to all orders.

\section{Conclusion\label{s:conclus}}

The main result of this paper is that $\gamma_{\rm str}^{(0)} =1/2$ for Nambu-Goto  bosonic strings in target-space dimensions
$2 < d<26$.  Our setup was designed to avoid any problem with tachyonic modes of the bosonic string. 
The origin of this half-integer value is the square-root dependence 
of the effective action on $J$ through $\blambda(J)$ given by \eq{newC} which is the true minimum.
The usual classical string ground state is
stable only for $d<2$, where zero-point fluctuations indeed increase  the effective action. In the formal limit $d < 2$ we thus 
expect the standard results from the Liouville theory, and it is seen explicitly from the formulas that the limit $d \to 2^+$ is somewhat 
singular.

The value
 $\gamma_{\rm str}^{(0}=1/2$ is clearly outside the  standard range of KPZ-DDK.
 We used explicitly
the Pauli-Villars regularization when calculating the effective action. 
However, it agrees with the value obtained for bosonic strings in $2 < d < 26$  
 using a  hyper-cubic lattice~\cite{lattice}.
 It is  interesting to understand if our effective bosonic  string theory, which has $\gamma_{\rm str}^{(0)}=1/2$, can still
 be viewed as a conformal invariant worldsheet theory.

\subsection*{Acknowledgment }

Y.M.\ was supported by the Russian Science Foundation (Grant No.20-12-00195).
Y.M.\ thanks the Theoretical Particle Physics and Cosmology group at NBI  for the 
long-term warm hospitality. 



\begin{thebibliography}{99}

\bibitem{KPZ} 
  V.G.~Knizhnik, A.M.~Polyakov and A.B.~Zamolodchikov,
  {\it Fractal structure of 2D quantum gravity,}
  Mod.\ Phys.\ Lett.\ A {\bf 3}, 819 (1988).\\
F.~David,
{\it Conformal field theories coupled to 2D Gravity in the conformal gauge,}
  Mod.\ Phys.\ Lett.\ A {\bf 3}, 1651 (1988).\\
  J.~Distler and H.~Kawai,
  {\it Conformal field theory and 2D quantum gravity,} 
  Nucl.\ Phys.\ B {\bf 321}, 509 (1989).

\bibitem{lattice}
B. Durhuus, J. Frohlich and  T. Jonsson, 
{\it Selfavoiding and planar random surfaces on the lattice},
Nucl.\ Phys.\  B {\bf 225},  185 (1983); 
{\it Critical behavior in a model of planar random surfaces},
Nucl.\  Phys.\  B {\bf 240},  453 (1984), Phys.\  Lett.\  B {\bf 137},  93 (1984).

\bibitem{ADJ97} 
  J.~Ambjorn, B.~Durhuus and T.~Jonsson,
 {\it Quantum geometry. A statistical field theory approach,}
  Cambridge  (UK) Univ.\ Press (1997).

  \bibitem{AM17a}
 J.~Ambjorn and Y.~Makeenko,	
{\it String theory as a Lilliputian world},
Phys.\ Lett.\ B {\bf 756}, 142  (2016)
[arXiv:1601.00540]; 
{\it Scaling behavior of regularized bosonic strings},
Phys.\ Rev.\ D\ {\bf 93},  066007 (2016) 
[arXiv:1510.03390];
{\it Stability of the nonperturbative bosonic string vacuum},
Phys.\ Lett.\ B {\bf 770}, 352 (2017)
[arXiv:1703.05382 [hep-th]].   

\bibitem{AM17c}
 J.~Ambjorn and Y.~Makeenko,	
{\it The use of Pauli-Villars' regularization in string theory,} 
 Int. J. Mod.\ Phys.\ A {\bf 32},  1750187 (2017)
 [arXiv:1709.00995 [hep-th]].

\bibitem{bergfin}
B.~Durhuus,
{\it Multispin systems on a randomly triangulated surface,}
Nucl. Phys. B \textbf{426}, 203    (1994)
[arXiv:9402052 [hep-th]].

\bibitem{Pol86} 
  J.~Polchinski,
  {\it Evaluation of the one loop string path integral,}
  Commun.\ Math.\ Phys.\  {\bf 104}, 37 (1986).

\bibitem{Zam82} 
  A.~B.~Zamolodchikov,
{\it On the entropy of random surfaces,}
  Phys.\ Lett.\  {\bf 117B}, 87 (1982).\\
  S.~Chaudhuri, H.~Kawai and S.~Tye,
{\it Path integral formulation of closed strings,}
  Phys.\ Rev.\ D {\bf 36}, 1148 (1987).\\
  I.~K.~Kostov and A.~Krzywicki,
{\it On the entropy of random surfaces with arbitrary genus,}
  Phys.\ Lett.\ B {\bf 187}, 149 (1987).
  



\end{thebibliography}
\end{document}